\documentclass[fleqn,usedcolumn,usenatbib]{mnras-fix}
\usepackage[british]{babel}             
\usepackage{newtxtext}                  
\usepackage[slantedGreek]{newtxmath}    
\usepackage[T1]{fontenc}                
\DeclareFontShape{OML}{cmm}{b}{it}{
       <5><6><7><8><9>gen*cmmib%
       <10><10.95><12><14.4><17.28><20.74><24.88>cmmib10%
       <5.5>cmmib5%
       <16>cmmib10}{}
\usepackage{graphicx}                   
\usepackage{booktabs}                   
\makeatletter\NAT@longnamestrue\makeatother
\newcommand{\dhead}[1]{\multicolumn{1}{c}{#1}}
\hypersetup{pdfauthor={D. A. Green},
  pdftitle={The integrated radio spectrum of G2.4+1.4},
  pdfkeywords={ISM: individual objects: G2.4+1.4, radio continuum: ISM,
     radiation mechanisms: thermal, radiation mechanisms: non-thermal},
  bookmarksnumbered=true}
\TheAuthor
\volume{{\rm in press}}
\title{The integrated radio spectrum of G2.4\texorpdfstring{$+$}{+}1.4}

\author[Green]{D.~A.~Green\thanks{email: {\tt dag@mrao.cam.ac.uk}}\\
 Astrophysics Group, Cavendish Laboratory, 19 J.~J.~Thomson Avenue,
 Cambridge CB3 0HE}

\date{Accepted 2022 August 16. Received 2022 August 15; in original form 2022 May 24}

\pubyear{2022}
\begin{document}

\label{firstpage}

\pagerange{\pageref{firstpage}--\pageref{lastpage}}

\maketitle

\begin{abstract}
The Galactic source G2.4$+$1.4 is an optical and radio nebula containing
an extreme Wolf--Rayet star. At one time this source was regarded as a
supernova remnant, because of its apparent non-thermal radio spectrum,
although this was based on limited observations. Subsequent observations
instead supported a flat, optically thin thermal radio spectrum for
G2.4$+$1.4, and it was identified as a photoionized, mass-loss bubble,
not a supernova remnant. Recently, however, it has been claimed that
this source has a non-thermal integrated radio spectrum. I discuss the
integrated radio flux densities available for G2.4$+$1.4 from a variety
of surveys, and show that it has a flat spectrum at gigahertz
frequencies (with a spectral index $\alpha$ of $0.02 \pm 0.08$,
where flux density $S$ scales with frequency $\nu$ as $S \propto
\nu^{-\alpha}$).
\end{abstract}


\section{Introduction}

G2.4$+$1.4 is an optical and radio source in Sagittarius, $\approx
12$~arcmin in extent (e.g.\ \citealt{1982ApJ...254..132T,
1987MNRAS.225..221G, 1995ApJ...439..637G}), showing multiple rings. The
Wolf--Rayet star WR~102 is within G2.4$+$1.4, but is offset from its
centre (e.g.\ \citealt{1990ApJ...351..563D}). It was identified as a
supernovae remnant (SNR) based on its non-thermal radio spectrum, as
reported by \citet{1975ApJ...198..111J}. Hence it was included in some
catalogues of Galactic supernova remnants (e.g.\
\citealt{1979AuJPh..32...83M, 1984MNRAS.209..449G}), although it was
removed from subsequent catalogues (i.e.\
\citealt{1988Ap+SS.148....3G}), as the non-thermal radio spectrum was
questioned by \citet{1987MNRAS.225..221G}. \citet{1990ApJ...359..419D}
identify G2.4$+$1.4 as a `photoionized, mass-loss bubble' powered by the
WR star. Recently, however, \citet{2019ApJ...884L..49P} have reported a
non-thermal integrated radio spectrum for G2.4$+$1.4. Here I investigate
the radio spectrum of G2.4$+$1.4 from available radio surveys, and
compare this with the results of \citet{2019ApJ...884L..49P}.

Section~\ref{s:background} provides background information on G2.4$+$1.4
and its radio spectrum. The radio spectrum of G2.4$+$1.4 based on
available survey observations is presented Section~\ref{s:spectrum}, and
this is discussed in Section~\ref{s:d+c}.

\section{Background}\label{s:background}

G2.4$+$1.4 appeared\footnote{In some cases labelled `G2.3$+$1.4'.} in
various single dish radio observations: i.e.\
\citet{1968ApJ...154L..75G} at 5000~MHz, \citet{1969AuJPA..11...27B} at
2650~MHz, \citet{1970A+AS....1..319A} at 2695~MHz, and
\citet{1971AuJPh..24..769S} at 1410~MHz, as listed in
\citet{1975ApJ...198..111J}. These single dish observations are of
low-resolution ($\approx 4$~arcmin at best), and do not resolve
G2.4$+$1.4 well, some giving sizes for the source between 5 and
9~arcmin. The flux densities between 1410 and 5000~MHz are between 2 and
5~Jy. \citet{1975ApJ...198..111J} presented higher frequency
observations of G2.4$+$1.4, at both 15.5 and 31.4~GHz, with much smaller
flux densities than the available flux densities at lower frequencies.
However, no error was given for the 15.5-GHz flux density, and the
31.4-GHz value was very uncertain ($0.34 \pm 0.26$~Jy, i.e.\ a nominal
detection at less than the 1.5$\sigma$ level). Because of the lower flux
densities at the higher frequencies \citeauthor{1975ApJ...198..111J}
deduced a non-thermal radio spectrum for G2.4$+$1.4, and identified it
as supernova remnant, albeit one that unusually contains an high
excitation star. Subsequently \citet{1982ApJ...254..132T} reported
optical observations of G2.4$+$1.4, showing a `complex double shell
structure'. They concluded that G2.4$+$1.4 is a SNR, and noted that its
non-thermal radio emission was the strongest argument for this
identification.

\begin{table*}
\caption{Radio flux densities for G2.4$+$1.4 from the literature, and
from S-PASS (see Appendix~\ref{s:appendix}).}\label{t:fluxes}
\begin{tabular}{d{0}d{3}d{2}d{2}ll}
                   &                    & \multicolumn{2}{c}{flux density} &                  &              \\
 \dhead{frequency} & \dhead{resolution} & \dhead{value}  & \dhead{error}   & telescope, notes & Reference(s) \\
 \dhead{/MHz}      & \dhead{/arcmin}    & \dhead{/Jy}    & \dhead{/Jy}     &                  &              \\\toprule
  1410     & 14    &  5     &        & Parkes 64~m                      & \citet{1975ApJ...198..111J}, from \citet{1971AuJPh..24..769S}                     \\
  2650     &  7.4  &  4     &        & Parkes 64~m                      & \citet{1975ApJ...198..111J}, from \citet{1969AuJPA..11...27B}                     \\
  2695     & 11.3  &  2     &  0.6   & NRAO 140~ft                      & \citet{1975ApJ...198..111J}, from \citet{1970A+AS....1..319A}                     \\
  5000     &  3.95 &  3.0   &  0.6   & Parkes 64~m                      & \citet{1975ApJ...198..111J}, from \citet{1968ApJ...154L..75G}                     \\
 15500     &  2.2  &  0.44  &        & Haystack 36.5~m                  & \citet{1975ApJ...198..111J}                                                       \\
 31400     &  3.75 &  0.32  &  0.26  & NRAO 11~m                        & \citet{1975ApJ...198..111J}                                                       \\[7pt]
  5000     &  4    &  2.5   &        & Parkes 64~m                      & \citet{1987A+A...171..261C}                                                       \\[7pt]
   843     &  1.1  &  2.61  &  0.20  & Molonglo, total flux density     & \citet{1994MNRAS.270..835G}                                                       \\
   843     &  1.1  &  2.55  &  0.22  & Molonglo, excluding SW component & \citet{1994MNRAS.270..835G}                                                       \\
  1408     &  9.4  &  3.46  &  0.35  & Effelsberg 100~m                 & \citet{1994MNRAS.270..835G}, from \citet{1990A+AS...83..539R}                     \\
  2695     &  4.3  &  2.20  &  0.44  & Effelsberg 100~m                 & \citet{1994MNRAS.270..835G}, from \citet{1984A+AS...58..197R}                     \\[7pt]
  1490     &  0.3  &  2.4   &  0.2   & VLA                              & \citet{1995ApJ...439..637G}                                                       \\[7pt]
%
  4850     &  4.2  &  3.55  &        & PMN survey, Parkes 64~m          & \citet{1997ApJ...488..224K}                                                       \\[7pt]
  8350     &  9.7  &  2.58  &  0.19  & GBES 13.7~m                      & \citet{2000AJ....119.2801L}                                                       \\[7pt]
   605     &  0.25 &  2.60  &  0.26  & uGMRT (band 4)                   & \citet{2019ApJ...884L..49P}                                                       \\
   640     &  0.25 &  2.54  &  0.25  & uGMRT (band 4)                   & \citet{2019ApJ...884L..49P}                                                       \\
   675     &  0.25 &  2.23  &  0.22  & uGMRT (band 4)                   & \citet{2019ApJ...884L..49P}                                                       \\
   710     &  0.25 &  2.17  &  0.22  & uGMRT (band 4)                   & \citet{2019ApJ...884L..49P}                                                       \\
   745     &  0.25 &  1.99  &  0.20  & uGMRT (band 4)                   & \citet{2019ApJ...884L..49P}                                                       \\
  1297     &  0.25 &  1.34  &  0.14  & uGMRT (band 5)                   & \citet{2019ApJ...884L..49P}                                                       \\
  1327     &  0.25 &  1.31  &  0.13  & uGMRT (band 5)                   & \citet{2019ApJ...884L..49P}                                                       \\
  1361     &  0.25 &  1.30  &  0.13  & uGMRT (band 5)                   & \citet{2019ApJ...884L..49P}                                                       \\
  1395     &  0.25 &  1.26  &  0.13  & uGMRT (band 5)                   & \citet{2019ApJ...884L..49P}                                                       \\
  1429     &  0.25 &  1.21  &  0.12  & uGMRT (band 5)                   & \citet{2019ApJ...884L..49P}                                                       \\[7pt]
  2303     &  8.9  &  3.29  &  0.49  & S-PASS, Parkes 64~m              & this paper, from \citet{2019MNRAS.489.2330C}                                      \\\bottomrule
\end{tabular}
\end{table*}

The non-thermal radio spectrum, and the SNR identification for
G2.4$+$1.4 was not supported by \citet{1987A+A...171..261C}, who
presented 5-GHz radio observations of the source. They derived a flux
density of 2.5~Jy for G2.4$+$1.4 at 5~GHz, with a size of $6 \times
8$~arcmin$^2$, and detected hydrogen recombination line emission from
it. Hence they concluded the radio emission from G2.4$+$1.4 is thermal.
\citet{1987MNRAS.225..221G} presented VLA observations of G2.4$+$1.4 at
4860~MHz with a much higher resolution ($\approx 31\times11$~arcsec$^2$)
than the previously available single dish radio observations. These show
a double ring/shell structure for G2.4$+$1.4, which correlates well with
the H$\alpha$ emission, but with no significant linear polarization,
which might be expected if the radio emission was non-thermal.
\citeauthor{1987MNRAS.225..221G} also questioned the non-thermal
spectrum reported by \citet{1975ApJ...198..111J}, and hence the SNR
identification for G2.4$+$1.4. The higher frequency flux densities
reported by \citet{1975ApJ...198..111J} are uncertain, and the flux
density at 31.4~GHz was based on only a single observation with a
$2$~arcmin beam, using a 6~arcmin uniform disc model. From the improved
resolution available from the VLA image of G2.4$+$1.4, the position of
the single 31.4-GHz observation ($17^{\rm h} 42^{\rm m} 43^{\rm s}$,
$-26^\circ 10\fm4$, B1950.0) lies in a region of low emission, and hence
the integrated flux density based on this, and the simple uniform disc
model -- which is smaller than the extent of G2.4$+$1.4 -- will
underestimate the true flux density. Furthermore, the limited area
observed by \citet{1975ApJ...198..111J} at 15.5~GHz may not be
sufficient to properly define the local background emission away from
G2.4$+$1.4, which would mean the integrated flux density would be an
underestimate.

\citet{1994MNRAS.270..835G} presented observations of G2.4$+$1.4 at
843~MHz, and derived a spectral index $\alpha =
0.13^{+0.26}_{-0.23}$ between 843 and 2695~MHz (here $\alpha$ is defined
such that flux density $S$ scales with frequency $\nu$ as $S \propto
\nu^{-\alpha}$), which did not support a non-thermal radio spectrum for
G2.4$+$1.4. The spectral index derived by
\citeauthor{1994MNRAS.270..835G} was based on the integrated 843~MHz
flux density, and results from more recent single dish surveys than
those available to \citeauthor{1975ApJ...198..111J} (see
Section~\ref{s:spectrum} for some further discussion).
\citet{1995ApJ...439..637G} presented further VLA observations
G2.4$+$1.4. These, at 1490~MHz, image the extended emission better than
the previous VLA observations at 4869~MHz by
\citeauthor{1987MNRAS.225..221G}. \citeauthor{1995ApJ...439..637G} also
conclude that G2.4$+$1.4 has a flat, thermal radio spectrum. Recently,
however, \citet{2019ApJ...884L..49P} have presented uGMRT observations
of G2.4$+$1.4 in `Band 4' (550--850~MHz) and Band 5 (1050--1450~MHz).
From these they derive a non-thermal spectral index of $\alpha = 0.83
\pm 0.10$, quite different from the approximately flat, thermal spectra
obtained by \citet{1994MNRAS.270..835G} and by
\citet{1995ApJ...439..637G}.

\begin{figure*}
\centerline{\includegraphics[angle=0,width=17.5cm]{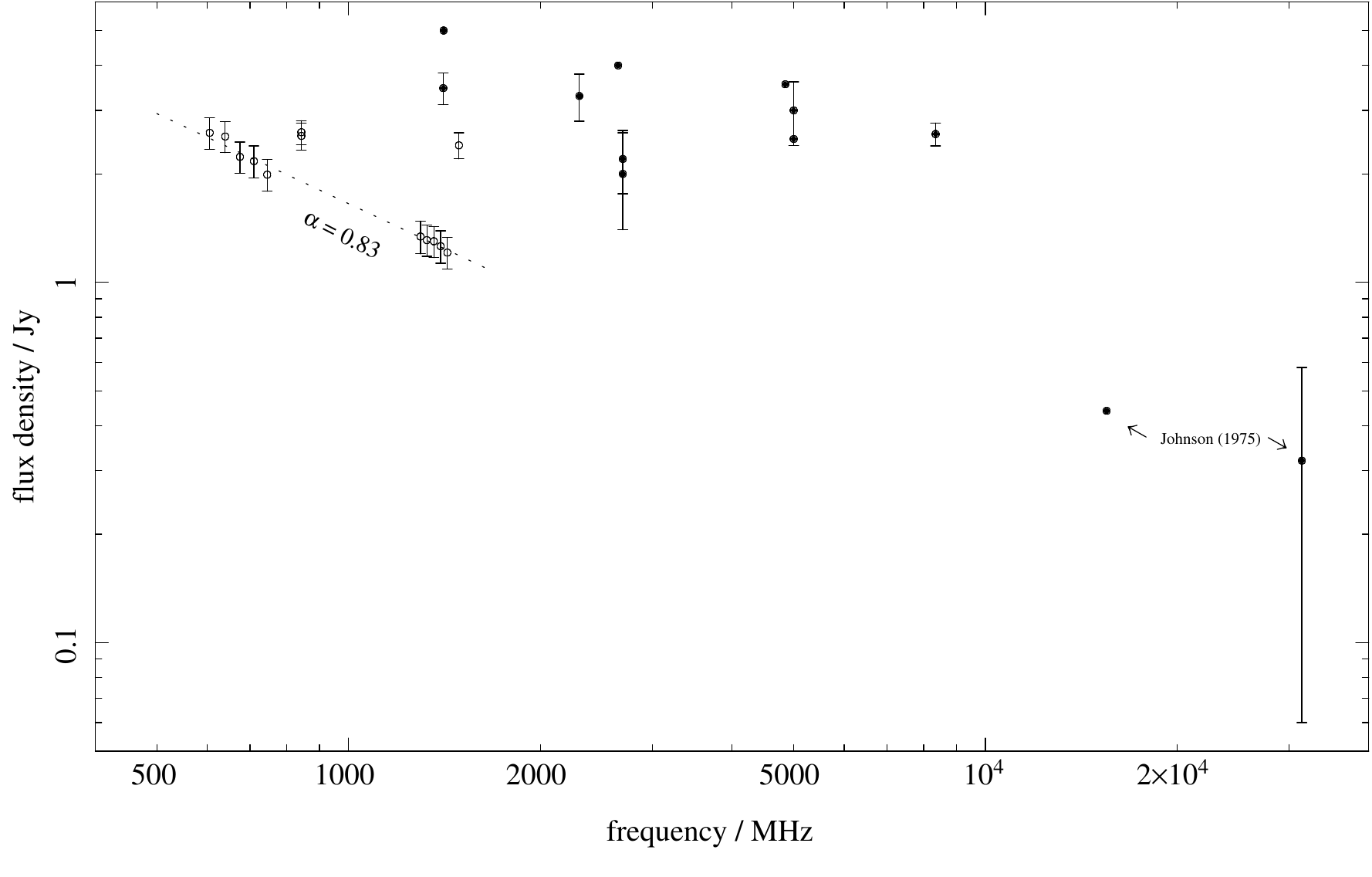}}
\caption{Radio spectrum of G2.4$+$1.4 using flux densities from the
literature, see Table~\ref{t:fluxes}. Filled circles are used for flux
densities from single dish observations, and open circles are used for
flux densities from interferometers, i.e.\ Molonglo at 843~MHz
\citep{1994MNRAS.270..835G}, VLA at 1490~MHz
\citep{1995ApJ...439..637G}, and uGMRT at 605 to 745~MHz and 1297 to
1429~MHz \citep{2019ApJ...884L..49P}. Error bars are included when flux
density errors are available in Table~\ref{t:fluxes}. The dotted line
shows a non-thermal spectrum with $\alpha = 0.83$ that
\citet{2019ApJ...884L..49P} obtained from their uGMRT flux
densities.}\label{f:spectrum}
\end{figure*}

%
\section{The radio spectrum of G2.4\texorpdfstring{$+$}{+}1.4}\label{s:spectrum}

Various radio flux densities for G2.4$+$1.4 from the literature cited in
Section~\ref{s:background}, and other sources, are given in
Table~\ref{t:fluxes} and plotted in Fig.~\ref{f:spectrum}. (The
resolution of the observations are also listed in Table~\ref{t:fluxes}.
For the Molonglo and VLA observations, at 843 and 1490~MHz
respectively, the geometrical mean of the major and minor axes of the
elliptical beam of the observations is given.) These flux densities
include a value at 2303~MHz derived from the S-PASS image, see
Appendix~\ref{s:appendix}.

At 843~MHz, two values are given from \citet{1994MNRAS.270..835G}, one
for the whole source, and the other excluding some emission to southwest
which was not seen at 4860~MHz in \citet{1987MNRAS.225..221G}. The
emission in the southwest, with $\approx 60$~mJy, was thought probably
to be due to an unrelated source with a steep spectrum.

G2.4$+$1.4 is included in the `Tropical' part of the Parkes--MIT--NRAO
(PMN) Survey at 4850~MHz, using observations made with the Parkes 64-m
telescope \citep{1994ApJS...90..179G}. The processing of the
PNM observations included the removal of a baseline, using a median
filter corresponding to a scale of 57~arcmin, which is several larger
than angular extent of G2.4$+$1.4. In the PMN catalogue the source has
a flux density of $1.027 \pm 0.054$~Jy, from a `fixed width' fit (see
\citealt{1993AJ....105.1666G}), including a flag indicating the sources
is `probably extended'. Given the resolution of the PMN survey is
$\approx 4.2$~arcmin then G2.4$+$1.4 is indeed expected to be resolved,
so an integrated flux density taking into account its extent is needed.
\citet{1997ApJ...488..224K} used the PMN survey images to obtain a peak
flux density of 1.025~Jy and angular size of 9.3~arcmin (based on a
Gaussian fit) for G2.4$+$1.4 from the PMN observations. This corresponds
to an integrated flux density of 3.55~Jy using the formula
provided\footnote{The integrated flux density is $S_{\rm p}
(\theta/\theta_{\rm psf})^2$, where $S_{\rm p}$ is the peak flux
density, $\theta$ is the fitted angular size, and $\theta_{\rm psf}$ is
the FWHM of the PMN images, 5~arcmin.} in \citet{1997ApJ...488..224K}.
This integrated flux density from \citet{1997ApJ...488..224K} is given
in Table~\ref{t:fluxes} and plotted in Fig.~\ref{f:spectrum} rather than
the flux density from the PNM catalogue.

In addition G2.4$+$1.4 is included in the region covered by the GPA
survey \citep{2000AJ....119.2801L}, which is at both 8.35 and 14.35~GHz.
It is in the source catalogue from this survey, but only at 8.35~GHz, as
`GPA G002.36+1.42' (the flux density limits for the two GPA frequencies
are 0.9~Jy at 8.35-GHz and 2.5~Jy at 14.35~GHz). This has a peak flux
density of 1.65~Jy, and an integrated flux density of 2.58~Jy, with a
size of $9.0 \times 5.2$~arcmin$^2$. The nominal uncertainty in
integrated flux density -- from both the accuracy of the flux density
scale (5.6~per~cent) and the rms noise on the images
(0.13~Jy~beam$^{-1}$) added in quadrature -- is 0.19~Jy.

\section{Discussion and Conclusions}\label{s:d+c}

Comparison of the different flux densities in Table~\ref{t:fluxes} is
not straightforward, given G2.4$+$1.4 is resolved, and the reported
fluxes densities have been integrated in different ways. For example,
consider the two flux densities from the Effelsberg 100-m surveys at
1408 and 2695~MHz of $3.46 (\pm 10~{\rm per~cent})$ and $2.20 (\pm
20~{\rm per~cent})$~Jy respectively, as listed by
\citet{1994MNRAS.270..835G}. These flux densities and errors give a
spectral index of $\alpha = 0.70$, but with a large error of $\pm 0.34$.
However, in the Effelsberg 100-m survey catalogues the sizes of the
G2.4$+$1.4 is given as $14.3 \times 14.0$~arcmin$^2$ at 1408~MHz and
$10.8 \times 8.2$~arcmin$^2$ at 2695~MHz. These sizes correspond to
deconvolved sizes $10.8 \times 10.4$~arcmin$^2$ and $9.9 \times
7.0$~arcmin$^2$, with the latter being 30~per~cent smaller in area,
which implies these integrated flux densities are not directly
comparable. Another issue is that observations made with interferometers
may not recover the total emission from an extended source. If the
observations lack sufficiently small baselines to adequately measure the
largest scales of the emission from a source, the flux densities will be
underestimates.
In this case it would be expected that, for a particular set of
baselines, this would be more significant at the higher frequencies for
observations made over a range of frequencies (as a fixed baseline
corresponds to a smaller angular scale at the higher frequencies).
This issue potentially applies to several flux densities
listed in Table~\ref{t:fluxes}: the 843~MHz Molonglo flux densities from
\citet{1994MNRAS.270..835G}; the 1490~MHz VLA flux density from
\citet{1995ApJ...439..637G}; the 605 to 1429~MHz uGMRT flux densities
from \citet{2019ApJ...884L..49P}.

If the higher frequency flux densities at 15.5 and 31.4~GHz from
\citet{1975ApJ...198..111J} are discounted -- see
Section~\ref{s:background} -- the available flux densities in
Table~\ref{t:fluxes} (see Fig.~\ref{f:spectrum}), apart from those from
\citeauthor{2019ApJ...884L..49P} are consistent with a flat radio
spectrum for G2.4$+$1.4. A weighted least-square fit to these flux
densities which have errors (i.e.\ those at 843 (the higher, total
value), 1408, 1490, 2303, 2695 (two values), 5000 and 8350~MHz in
Table~\ref{t:fluxes}) gives a spectral index $\alpha = 0.02 \pm 0.08$.
For this fit the errors in these flux densities were taken as double the
values listed in Table~\ref{t:fluxes} ,to be cautious (given that flux
densities have been derived in different ways, as noted above). This
flat spectrum is not consistent with the non-thermal spectrum with
$\alpha = 0.83 \pm 0.10$ derived by \citeauthor{2019ApJ...884L..49P}, as
is seen in Fig.~\ref{f:spectrum}. The uGMRT Band~5 flux densities
(i.e.\ 1297 to 1429~MHz) are significantly lower than the integrated
flux densities from the Molonglo observations at 843~MHz and the several
single dish surveys at $\approx 1.4$ to $8.35$~GHz. This discrepancy is
larger at the higher frequencies within the uGMRT band.

The images shown by \citeauthor{2019ApJ...884L..49P} do not show as much
extended emission as is seen in the VLA image by
\citet{1995ApJ...439..637G}. The VLA observations were made in
D-configuration, which has better smaller baseline coverage than the
uGMRT. This implies that the uGMRT images are missing larger scale
emission, and hence will underestimate the flux density of G2.4$+$1.4.
Moreover, the uGMRT images show less extended emission at the higher
frequencies. \citeauthor{2019ApJ...884L..49P} used a fixed radius range
in the $uv$-plane, which might be expected to eliminate any problem with
missing large scale structure varying with frequency. However, the
minimum radius used by \citeauthor{2019ApJ...884L..49P} was small,
corresponding to only 35~m at 1429~MHz. The uGMRT does not have good
coverage in the $uv$-plane at this radius (the smallest separations
between antennas is $\approx 103$~m). Also, at the higher uGMRT
frequencies, this minimum $uv$-radius is less than the 45~m diameter of
the uGMRT antennas, so observations from such small baselines will
suffer from shadowing between antennas. Moreover,
\citeauthor{2019ApJ...884L..49P} calculate their integrated flux
densities only for emission greater than a $3\sigma$ level, which
produces biased results that depend systematically on the $\sigma$
value.

Given the flux densities for G2.4$+$1.4 that are available from various
surveys, it has a flat, optically thin thermal radio spectrum at
gigahertz frequencies. The steep, non-thermal integrated radio spectrum
reported by \citet{2019ApJ...884L..49P} for G2.4$+$1.4 is not consistent
with the flat spectrum from other available observations.

Problems deriving accurate radio spectral indices for extended sources
from interferometer observations, which miss large-scale structure, do
not appear to be limited to the case of G2.4$+$1.4. There are several
reports, based on VLA observations, of non-thermal radio emission from
stellar bow-shocks \citep{2010A+A...517L..10B, 2021MNRAS.503.2514B,
2022A+A...663A..80M}. However, there is extended emission in the region
of these sources on scales larger than can be measured by the VLA
configuration used (an issue that is worse at the higher frequencies).
Hence there are significant systematic uncertainties in the apparent
non-thermal spectral indices that have been reported for these sources.

\section*{Acknowledgements}

This research has made use of NASA's Astrophysics Data System, and
S-band Polarisation All Sky Survey (S-PASS) data.

\section*{Data Availability}

The data underlying this article are available in the article, or -- for
S-PASS -- from \url{https://sites.google.com/inaf.it/spass}.

%
\phantomsection

\appendix
%
%
\section[A 2303-MHz flux density from S-PASS]{A 2303-MH\lowercase{z}
flux density from S-PASS}\label{s:appendix}

\begin{figure}
\centerline{\includegraphics[angle=0,width=8.5cm]{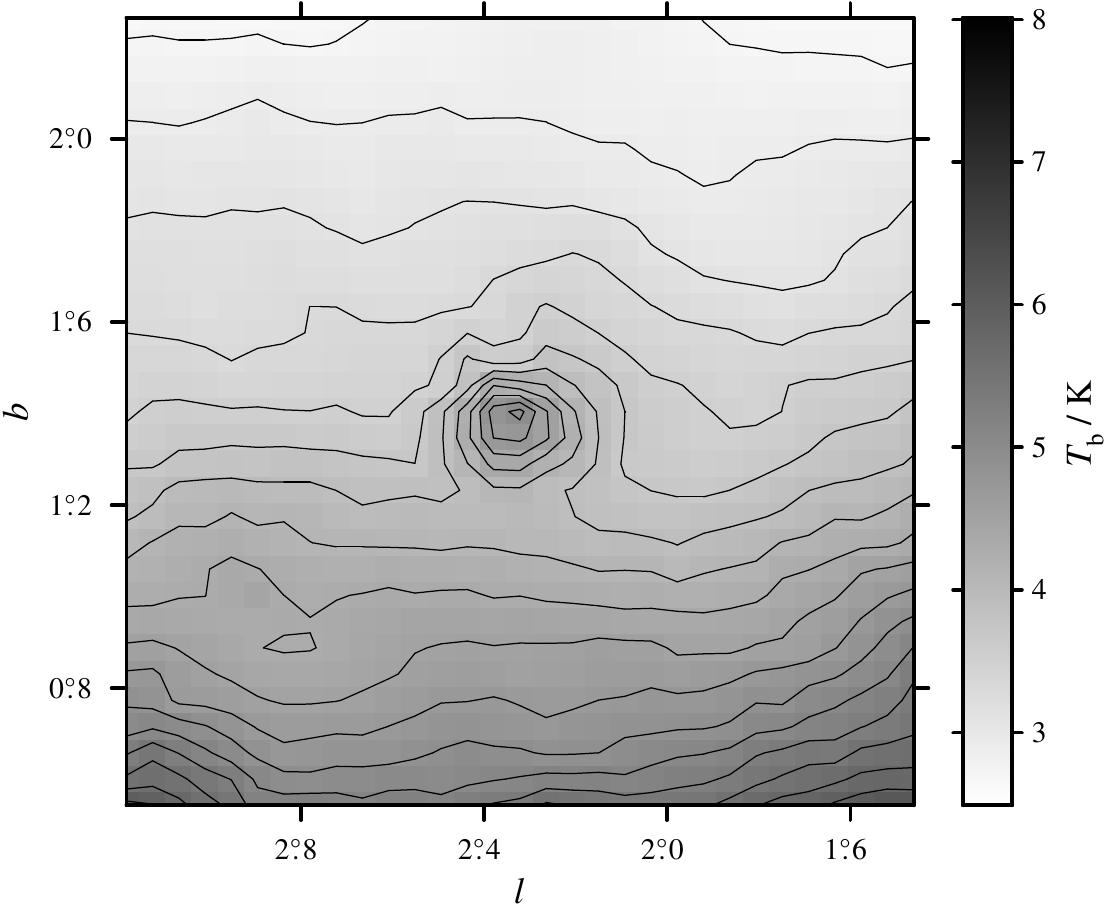}}
\caption{S-PASS image of G2.4$+$1.4, with Galactic coordinate labels.
The contour levels are every 0.2~K in brightness temperature, from 2.7
to 6.1~K, and the greyscale is from 2.0 to 8.0~K.}\label{f:spass-raw}
\end{figure}

\begin{figure}
\centerline{\includegraphics[angle=0,width=8.5cm]{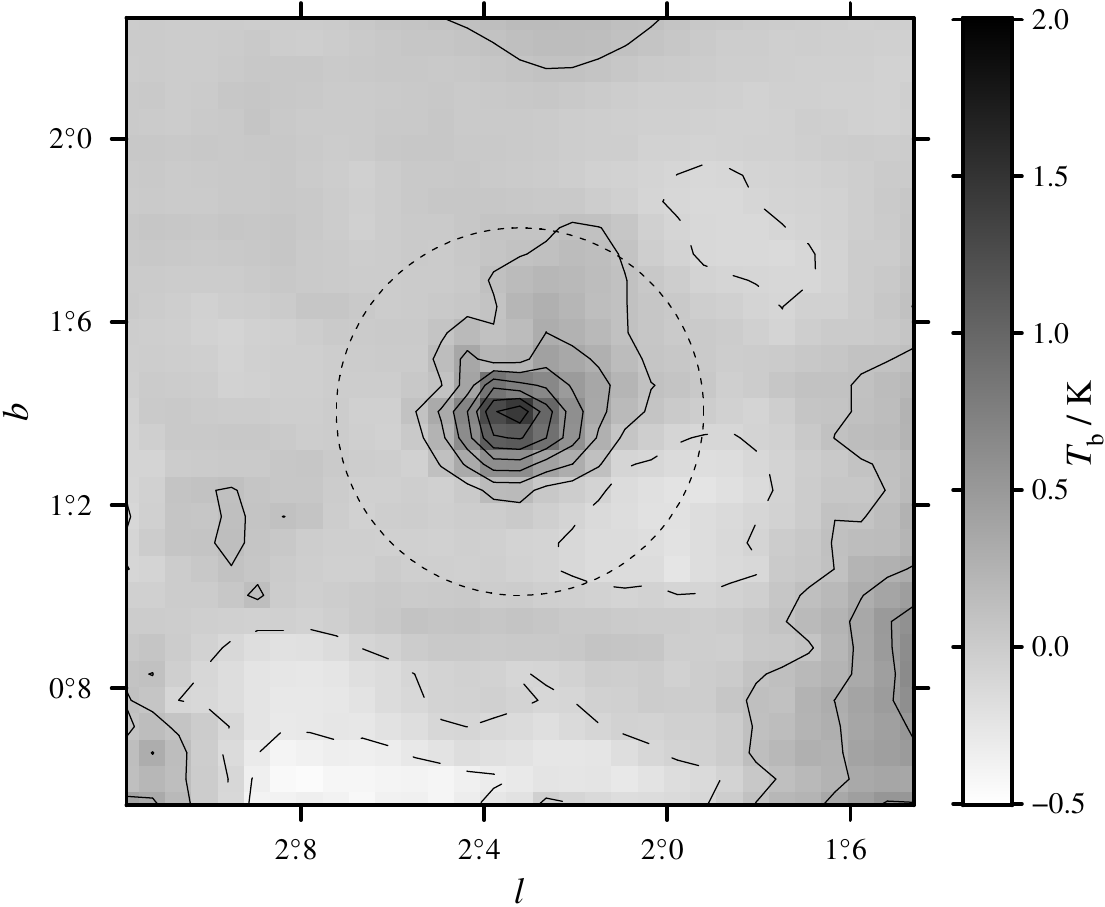}}
\caption{S-PASS image of G2.4$+$1.4, with Galactic coordinate labels,
after latitude dependant background removal (see text). The contours are
$\pm 0.3$, $\pm 0.1, 0.1, 0.3, \ldots, 1.3$~K in brightness temperature,
with negative contours dashed, and the greyscale is from $-0.5$ to
2.0~K. The dotted circle indicates the area integrated for the flux
density of G2.4$+$1.4.}\label{f:spass-rem}
\end{figure}

G2.4$+$1.4 is included in the region covered by the S-band Polarisation
All Sky Survey (S-PASS), see \citet{2019MNRAS.489.2330C}. The S-PASS
observations were made at 2303~MHz with the Parkes 64-m telescope, with
a resolution of 8.9~arcmin. As the source catalogue from S-PASS
\citep{2017PASA...34...13M} excludes Galactic latitudes with $|b| <
10^\circ$, an integrated flux density for G2.4$+$1.4 is derived here
from the S-PASS image\footnote{See:
\url{https://sites.google.com/inaf.it/spass}.}. Figure~\ref{f:spass-raw}
shows the S-PASS image of G2.4$+$1.4 and its surroundings, which clearly
shows G2.4$+$1.4, with a peak of $\approx 5$~K. This is $\approx 1.5$~K
above the local background level, which corresponds to $\approx
1.8$~Jy~beam$^{-1}$. However, as G2.4$+$1.4 is slightly resolved in
S-PASS, an integrated flux density is required, which needs to take into
account the local background emission (which is at about twice the level
of the peak excess seen for G2.4$+$1.4). In this case the background
gradient is larger at lower Galactic latitudes than at higher latitudes.
Hence simply averaging the local emission outside G2.4$+$1.4 in an
annulus would provide a biased estimate of local background. However,
from Fig.~\ref{f:spass-raw} is it evident that the variation in the
background emission is largely with Galactic latitude. The background
emission at a particular latitude, i.e.\ each row in
Fig.~\ref{f:spass-raw}, was estimated by taking the median of selected
pixel values in the row. The selected pixels excluded the central third
of each row, to exclude the emission from G2.4$+$1.4 (i.e.\ the regions
with Galactic longitude from approximately $1\fdg5$ to $2\fdg0$ and
$2\fdg6$ to $3\fdg2$ were used). Figure~\ref{f:spass-rem} shows the
S-PASS image after removal of this latitude dependant background
estimate. The integrated emission from G2.4$+$1.4 in
Fig.~\ref{f:spass-rem}, within the circle shown, is 3.26~Jy. The radius
of the circle is 7 pixels (each 3.44~arcmin), i.e.\ about 24~arcmin.
This radius is the approximate radius of G2.4$+$1.4 seen in higher
resolution images (i.e.\ $\approx 6$~arcmin), plus twice the S-PASS
resolution. The integrated flux density derived varies if the background
removal uses different ranges of Galactic longitude, or if a different
integration radius is chosen. For example, if the integration radius of
is changed by $1$~pixel then the integrated flux density changes by
4~per~cent. To be cautious I assign an considerably larger error of
15~per~cent to this integrated flux density (which is larger than the
flux density scale uncertainty of 5~per~cent for S-PASS), i.e.\ 0.49~Jy.

\bsp

\label{lastpage}
\end{document}